\documentclass[article]{jss}
\usepackage[utf8]{inputenc}

\pdfoutput=1

\author{
Jared Huling\\Department of Statistics\\
University of Wisconsin-Madison \And Peter Z.G. Qian\\Department of Statistics\\
University of Wisconsin-Madison
}
\title{Fast Penalized Regression and Cross Validation for Tall Data with the
\pkg{oem} Package}

\Abstract{
A large body of research has focused on theory and computation for
variable selection techniques for high dimensional data. There has
been substantially less work in the big ``tall'' data paradigm, where
the number of variables may be large, but the number of observations is
much larger. The orthogonalizing expectation maximization (OEM)
algorithm is one approach for computation of penalized models which
excels in the big tall data regime. The \pkg{oem} package is an
efficient implementation of the OEM algorithm which provides a multitude
of computation routines with a focus on big tall data, such as a function for out-of-memory
computation, for large-scale parallel computation of penalized regression models. Furthermore, in this paper we propose a specialized
implementation of the OEM algorithm for cross validation, dramatically
reducing the computing time for cross validation over a naive
implementation.
}

\Plainauthor{Jared Huling, Peter Z.G. Qian}
\Plaintitle{Fast Penalized Regression and Cross Validation for Tall Data with the
oem Package}
\Shorttitle{\pkg{oem}: A Package for Fast Penalized Regression}
\Plainkeywords{lasso, mcp, optimization, expectation maximization, C++, OpenMP, parallel computing, out-of-memory computing}

\Submitdate{}

\Address{
    Jared Huling\\
  Department of Statistics\\
  University of Wisconsin-Madison\\
  Wisconsin 53706, United States of America\\
  E-mail: \href{mailto:huling@wisc.edu}{\nolinkurl{huling@wisc.edu}}\\
  URL: \url{http://www.stat.wisc.edu/~huling/}\\~\\
    }

\usepackage{amsmath} \usepackage{longtable} \usepackage{booktabs}
\usepackage{bbm, bm}  \def\m{\boldsymbol}

\def\coef{\boldsymbol \beta} \def\hcoef{\hat{\boldsymbol \beta}}

\def\coef{\boldsymbol \beta} \def\hcoef{\hat{\boldsymbol \beta}}

\def\bfy{\mathbf y}  
\def\bfu{\mathbf u} \def\bfI{\mathbf I} \def\bfX{\mathbf X} \def\bfv{\mathbf v}
 \def\bfz{\mathbf z} 
 \def\bfu{\mathbf u} 
  
 \def\R{\mathbbm R} 
 \def\bfI{\mathbf I} 
 \def\argmin{\operatornamewithlimits{argmin}}
 \usepackage{algorithm}
\usepackage{algorithmic} \usepackage{float}

\begin{document}

\section[Introduction]{Introduction}\label{introduction}

Penalized regression has been a widely used technique for variable
selection for many decades. Computation for such models has been a
challenge due to the non-smooth properties of penalties such as the
lasso \citep{tibshirani96}. A plethora of algorithms now exist such as
the Least Angle Regression (LARS) algorithm \citep{efron2004} and
coordinate descent \citep{Tseng2001, friedman2010}, among many others.
There exist for these algorithms even more packages for various penalized 
regression routines such as
\pkg{glmnet} \citep{glmnet}, \pkg{lars} \citep{lars}, \pkg{ncvreg}
\citep{ncvreg}, \pkg{grpreg} \citep{grpreg}, and \pkg{gglasso}
\citep{gglasso}, among countless others. Each of the above packages
focuses on a narrow class of penalties, such as group regularization,
non-convex penalties, or the lasso. The existence of many options often
makes it hard to choose between packages, and furthermore difficult to
develop a consistent workflow when it is not clear which type of penalty
is suitable for a given problem. There has been much focus on algorithms for
scenarios where the number of variables \(p\) is much larger than the
number of observations \(n\), such as the LARS algorithm, yet many
applications of ``internet scale'' typically involve an extraordinarily
large number of observations and a moderate number of variables. In
these applications, speed is crucial. The \pkg{oem} package is intended
to provide a highly efficient framework for penalized regression in
these tall data settings. It provides computation for a comprehensive
selection of penalties, including the lasso and elastic net, group-wise
penalties, and non-convex penalties and allows for simultaneous
computation of these penalties. Most of the algorithms and packages
listed above, however, are more efficient than the \pkg{oem} package for
scenarios when the number of variables is larger than the number of
observations. Roughly speaking, \pkg{oem} package is most effective when
the ratio of the number of variables to the number of observations is
less than \(1/10\). This is ideal for data settings, such as in internet
applications, where a large number of variables are available, but the
number of observations grows rapidly over time.

Centered around the orthogonalizing expectation maximization (OEM)
algorithm of \citet{xiong16}, the \pkg{oem} package provides a unified
framework for computation for penalized regression problems with a focus
on big \emph{tall} data scenarios. The OEM algorithm is particularly
useful for regression scenarios when the number of observations is
significantly larger than the number of variables. It is efficient even
when the number of variables is large (in the thousands) as long as the
number of observations is yet larger (hundreds of thousands or more).
The OEM algorithm is particularly well-suited for penalized linear
regression scenarios when the practitioner must choose between a large
number of potential penalties, as the OEM algorithm can compute full
tuning parameter paths for multiple penalties nearly as efficiently as
for just one penalty.

The \pkg{oem} package also places an explicit emphasis on practical
aspects of penalized regression model fitting, such as tuning parameter
selection, in contrast to the vast majority of penalized regression
packages. The most common approach for tuning parameter selection for
penalized regression is cross validation. Cross validation is
computationally demanding, yet there has been little focus on efficient 
implementations of cross validation. Here we present a modification of the
OEM algorithm to dramatically reduce the computational load for cross
validation. Additionally, the \pkg{oem} package provides some extra
unique features for very large-scale problems. The \pkg{oem} package
provides functionality for out-of-memory computation, allowing for
fitting penalized regression models on data which are too large to fit
into memory. Feasibly, one could use these routines
to fit models on datasets hundreds of gigabytes in size on just a
laptop. The \pkg{biglasso} package \citep{biglasso} also provides
functionality for out-of-memory computation, however its emphasis is on
ultrahigh-dimensional data scenarios and is limited to the lasso,
elastic-net, and ridge penalties.  Also provided are OEM routines based off of the quantities
\(\bfX^\top\bfX\) and \(\bfX^\top\bfy\) that may already be available to
researchers from exploratory analyses. This can be especially useful for
scenarios when data are stored across a large cluster, yet the
sufficient quantities can be computed easily on the cluster, making
penalized regression computation very simple and quick for datasets with
an arbitrarily large number of observations.

The core computation for the \pkg{oem} package is in \proglang{C++}
using the \pkg{Eigen} numerical linear algebra library \citep{eigenweb}
with an \proglang{R} interface via the \pkg{RcppEigen} \citep{RcppEigen}
package. Out-of-memory computation capability is provided by interfacing
to special \proglang{C++} objects for referencing objects stored on disk
using the \pkg{bigmemory} package \citep{bigmemory}.

In Section \ref{the-orthogonalizing-em-algorithm}, we provide a review
of the OEM algorithm. In Section
\ref{parallelization-and-fast-cross-validation} we present a new
efficient approach for cross validation based on the OEM algorithm. In
Section \ref{extension-to-logistic-regression} we show how the OEM
algorithm can be extended to logistic regression using a proximal Newton
algorithm. Section \ref{the-oem-package} provides an introduction to the
package, highlighting useful features. Finally, Section \ref{timings}
demonstrates the computational efficiency of the package with some
numerical examples.

\section[The orthogonalizing EM
algorithm]{The orthogonalizing EM
algorithm}\label{the-orthogonalizing-em-algorithm}

\subsection[Review of the OEM
algorithm]{Review of the OEM
algorithm}\label{review-of-the-oem-algorithm}

The OEM algorithm is centered around the linear regression model:

\begin{equation}\label{linear_model}
\bfy=\bfX\coef+\boldsymbol{\varepsilon},
\end{equation}

where \(\bfX = (x_{ij})\) is an \(n \times p\) design matrix,
\(\bfy \in \R^n\) is a vector of responses,
\(\coef = (\beta_1,\ldots,\beta_p)^\top\) is a vector of regression
coefficients, and \(\boldsymbol{\varepsilon}\) is a vector of random
error terms with mean zero. When the number of covariates is large,
researchers often want or need to perform variable selection to reduce
variability or to select important covariates. A sparse estimate
\(\hat{\coef}\) with some estimated components exactly zero can be
obtained by minimizing a penalized least squares criterion:

\begin{equation}\label{eqn:pen_loss}
\hcoef=\argmin_{\coef}\|\bfy-\bfX\coef\|^2 + P_\lambda(\coef),
\end{equation}

where the penalty term \(P_\lambda\) has a singularity at the zero point
of its argument. Widely used examples of penalties include the lasso
\citep{tibshirani96}, the group lasso \citep{yuan2006}, the smoothly
clipped absolute deviation (SCAD) penalty \citep{fan01}, and the minimax
concave penalty (MCP) \citep{zhang2010}, among many others.

The OEM algorithm can solve (\ref{eqn:pen_loss}) under a broad class
of penalties, including all of those previously mentioned. The basic
motivation of the OEM algorithm is the triviality of minimizing this
loss when the design matrix \(\bfX\) is orthogonal. However, the
majority of design matrices from observational data are not orthogonal.
Instead, we seek to augment the design matrix with extra rows such that
the augmented matrix is orthogonal. If the non-existent responses of the
augmented rows are treated as missing, then we can embed our original
minimizaton problem inside a missing data problem and use the EM
algorithm. Let \(\boldsymbol{\Delta}\) be a matrix of pseudo
observations whose response values \(\bfz\) are missing. If
\(\boldsymbol{\Delta}\) is designed such that the augmented regression
matrix \[\bfX_c = \begin{pmatrix}\bfX \\ {\bf \Delta}\end{pmatrix}\] is
column orthogonal, an EM algorithm can be used to solve the augmented
problem efficiently similar to \cite{healy1956}. The OEM algorithm
achieves this with two steps:

\begin{description}
    \item[Step 1.] Construct an augmentation matrix ${\bf \Delta}$. 
    \item[Step 2.] Iteratively solve the orthogonal design with missing data by EM algorithm.
    \begin{description}
        \item[Step 2.1.] E-step: impute the missing responses $\bfz$ by $\bfz = \boldsymbol{\Delta}\coef^{(t)}$, where $\coef^{(t)}$ is the current estimate.
        \item[Step 2.2.] M-step: solve 
        \begin{equation} \label{mstep}
        \coef^{(t+1)} = \argmin_{\coef} \frac{1}{2}\|\bfy - \bfX\coef\|^2 + \frac{1}{2}\|\bfz - \boldsymbol{\Delta}\coef\|^2 + P_\lambda(\coef).
        \end{equation}
    \end{description}
\end{description}

An augmentation matrix \({\bf \Delta}\) can be constructed using the
active orthogonalization procedure. The procedure starts with any
positive definite diagonal matrix \(\boldsymbol{S}\) and
\({\bf \Delta}\) can be constructed conceptually by ensuring that
\({\bf \Delta}^\top{\bf \Delta} = d\boldsymbol{S}^2 - \bfX^\top\bfX\) is
positive semidefinite for some constant
\(d \geq \lambda_1(\boldsymbol{S}^{-1}\bfX^\top\bfX\boldsymbol{S}^{-1})\),
where \(\lambda_1(\cdot)\) is the largest eigenvalue. The term
\({\bf \Delta}\) need not be explicitly computed, as the EM iterations
in the second step result in closed-form solutions which only depend on
\({\bf \Delta}^\top{\bf \Delta}\). To be more specific, let
\({\bf A}={\bf \Delta}^\top{\boldsymbol{\Delta}}\) and
\({\bf u}=\bfX^\top\bfy+{\bf A}\coef^{(t)}\). Furthermore, assume the
regression matrix \(\m{X}\) is standardized so that

\begin{equation}
\sum_{i=1}^nx_{ij}^2=1, \;  \mbox{for} \; j=1,\ldots,p.\nonumber
\end{equation}

Then the update for the regression coefficients when
\(P_\lambda(\coef) = 0\) has the form \(\beta_j^{(t+1)}=u_j/d_j\). When
\(P_\lambda(\coef)\) is the \(\ell_1\) norm, corresponding to the lasso
penalty,

\begin{equation}
\beta_j^{(t+1)}={\mathrm{sign}}(u_j)\left(\frac{|u_j|-\lambda}{d_j}\right)_+, \nonumber
\end{equation}

where \((a)_+\) denotes \(\max\{a,0\}\).

\citet{xiong16} used a scalar value for \(d\) and for simplicity chose
\(S\) to be the diagonal matrix. Furthermore, they showed that while the
above algorithm converges for all \(d \geq \lambda_1(\bfX^\top\bfX)\),
using \(d = \lambda_1(\bfX^\top\bfX)\) results in the fastest
convergence. Such a \(d\) can be computed efficiently using the Lanczos
algorithm \citep{lanczos1950}.

\subsection[Penalties]{Penalties}\label{penalties}

The \pkg{oem} package uses the OEM algorithm to solve penalized least
squares problems with the penalties outlined in Table \ref{tab:pen}. For a vector ${\boldsymbol u}$ of length $k$ and an index set $g \subseteq \{ 1, \dots, k  \}$ we define the length $|g|$ subvector ${\boldsymbol u}_g$ of a vector ${\boldsymbol u}$ as the elements of ${\boldsymbol u}$ indexed by $g$. Furthermore, for a vector $\boldsymbol u$ of length $k$ let $|| \boldsymbol u|| = \sqrt{\sum_{j=1}^k{\boldsymbol u}_j^2}$.

\newcommand{\ra}[1]{\renewcommand{\arraystretch}{#1}}

\begin{table*}[h]
\centering
\ra{1.3}
\begin{tabular}{@{}ccc@{}}\toprule
 Penalty & Penalty form \\ \midrule
 Lasso & $\lambda \sum_{j = 1}^pw_j|\beta_j|$  \\
 Elastic Net & $\alpha\lambda \sum_{j = 1}^pw_j|\beta_j| + \frac{1}{2}(1 - \alpha)\lambda \sum_{j = 1}^pw_j\beta_j^2$  \\
 MCP & $\sum_{j = 1}^p P^{MCP}_{\lambda w_j,\gamma}(\beta_j)$  \\
 SCAD & $\sum_{j = 1}^p P^{SCAD}_{\lambda w_j,\gamma}(\beta_j)$  \\
 Group Lasso & $\lambda \sum_{k = 1}^Gc_k|| \coef_{g_k} ||$ \\
 Group MCP & $\lambda \sum_{k = 1}^G  P^{MCP}_{\lambda c_k,\gamma}(||\coef_{g_k}||) $ \\
 Group SCAD & $\lambda \sum_{k = 1}^G  P^{SCAD}_{\lambda c_k,\gamma}(||\coef_{g_k}||) $ \\
 Sparse Group Lasso & $\lambda(1 - \tau) \sum_{k = 1}^Gc_k|| \coef_{g_k} ||  + \lambda \tau \sum_{j = 1}^pw_j|\beta_j|$\\
\bottomrule
\end{tabular}
\caption{Listed above are the penalties available in the \pkg{oem} package. In the group lasso, ${g_k}$ refers to the index set of the $k$th group. The vector $\boldsymbol w \in \R^p$ is a set of variable-specific penalty weights and the vector $\boldsymbol c \in \R^G$ is a set of group-specific penalty weights.}
\label{tab:pen}
\end{table*}

For $\lambda > 0$ let:
 \[ P_{\lambda, \gamma}^{SCAD}(\beta) =   \left\{
\begin{array}{ll}
      \lambda|\beta| & |\beta| \leq \lambda ; \\
      -\frac{|\beta|^2 - 2\gamma\lambda|\beta| + \lambda^2}{2(\gamma - 1)} & \lambda < |\beta| \leq \gamma\lambda ; \\
      \frac{(\gamma + 1)\lambda^2}{2} & |\beta| > \gamma\lambda \\
\end{array} 
\right. \]
for $\gamma > 2$
and \[ P_{\lambda, \gamma}^{MCP}(\beta) =   \left\{
\begin{array}{ll}
      \lambda|\beta| - \frac{\beta^2}{2\gamma} & |\beta| \leq \gamma\lambda ; \\
      \frac{\gamma\lambda^2}{2} & |\beta| > \gamma\lambda \\
\end{array} 
\right. \]
for $\gamma > 1$.
 The updates for the above penalties are given below:

\begin{description}

\item[]1. \textbf{Lasso}
\begin{equation}\label{lasso}
\beta_j^{(t+1)}= S(u_j, w_j\lambda, d) = {\mathrm{sign}}(u_j)\left(\frac{|u_j|-w_j\lambda}{d}\right)_+.
\end{equation}

\item[]2. \textbf{Elastic Net}
\begin{equation}
\beta_j^{(t+1)}= {\mathrm{sign}}(u_j)\left(\frac{|u_j|-w_j\alpha\lambda}{d+w_j(1-\alpha)\lambda}\right)_+.\label{net}
\end{equation}

\item[]3. \textbf{MCP}
\begin{equation}
\beta_j^{(t+1)} = M(u_j, w_j\lambda, \gamma, d) =\left\{\begin{array}{ll}{\mathrm{sign}}(u_j)\frac{\gamma\big(|u_j|-w_j\lambda\big)_+}{(\gamma d-1)},\quad&\text{if}\
|u_j|\leq w_j\gamma\lambda d,
\\ u_j/d,&\text{if}\ |u_j|>w_j\gamma\lambda d.\end{array}\right.\label{mcp}
\end{equation}
where $\gamma > 1$

\item[]4. \textbf{SCAD}
\begin{equation}\label{scad} 
\beta_j^{(t+1)} = C(u_j, w_j\lambda, \gamma, d) =\left\{\begin{array}{ll}{\mathrm{sign}}(u_j)\big(|u_j|-w_j\lambda\big)_+/d,&\text{if}\
|u_j|\leq(d+1)w_j\lambda,
\\{\mathrm{sign}}(u_j)\frac{\big[(\gamma-1)|u_j|-w_j\gamma\lambda\big]}{\big[(\gamma-1)d-1\big]},\quad&\text{if}\ (d+1)w_j\lambda<|u_j|\leq w_j\gamma\lambda d,
\\ u_j/d,&\text{if}\ |u_j|>w_j\gamma\lambda d.\end{array}\right.
\end{equation}
where $\gamma > 2$.

\item[]5. \textbf{Group Lasso}

The update for the $k$th group is 
\begin{equation}\label{glasso}
\coef_{g_k}^{(t+1)} = G(\bfu_{g_k}, c_k\lambda, d) =\frac{\bfu_{g_k}}{d}\left(1 - \frac{c_k\lambda}{||\bfu_{g_k}||_2}\right)_+.
\end{equation}

\item[]6. \textbf{Group MCP}

The update for the $k$th group is 
\begin{equation}\label{gmcp}
\coef_{g_k}^{(t+1)} =  M(||\bfu_{g_k}||, c_k\lambda, \gamma, d)\frac{\bfu_{g_k}}{||\bfu_{g_k}||}.
\end{equation}

\item[]7. \textbf{Group SCAD}

The update for the $k$th group is 
\begin{equation}\label{gscad}
\coef_{g_k}^{(t+1)} = C(||\bfu_{g_k}||, c_k\lambda, \gamma, d)\frac{\bfu_{g_k}}{||\bfu_{g_k}||}.
\end{equation}

\item[]8. \textbf{Sparse Group Lasso}

The update for the $k$th group is 
\begin{equation}\label{sglasso}
\coef_{g_k}^{(t+1)} = G(\bfv_{g_k}, c_k\lambda(1 - \tau), 1),
\end{equation}
where the $j$th element of $\bfv$ is $S(u_j, w_j\lambda\tau, d)$. This is true because the thresholding operator of two nested penalties is the composition of the thresholding operators where innermost nested penalty's thresholding operator is evaluated first. For more details and theoretical justification of the composition of proximal operators for nested penalties, see \citet{jenatton2010proximal}.

\end{description}

\section[Parallelization and fast cross
validation]{Parallelization and fast cross
validation}\label{parallelization-and-fast-cross-validation}

The OEM algorithm lends itself naturally to efficient computation for
cross validation for penalized linear regression models. When the number
of variables is not too large (ideally \(n >> p\)) relative to the
number of observations, computation for cross validation using the OEM
algorithm is on a similar order of computational complexity as fitting
one model for the full data. To see why this is the case, note that the
key computational step in OEM is in forming the matrix \(\bf A\). Recall
that \(\bf A = d\bfI_p - \bfX^\top\bfX\). In \(K\)-fold cross
validation, the design matrix is randomly partitioned into \(K\)
submatrices as \[
\bfX = 
  \begin{pmatrix} \bfX_1 \\ \vdots \\ \bfX_K \end{pmatrix}. 
\] Then for the \(k^{th}\) cross validation model, the oem algorithm
requires the quantities
\(\bf A_{(k)} = d_{(k)}\bfI_p - \bfX_{-k}^\top\bfX_{-k}\) and
\(\bfX_{-k}^\top\bfy_{-k}\), where \(\bfX_{-k}\) is the design matrix
\(X\) with the \(k^{th}\) submatrix \(\bfX_{k}\) removed and
\(\bfy_{-k}\) is the response vector with the elements from the
\(k^{th}\) fold removed. Then trivially, we have that
\(\bf A = d\bfI_p - \sum_{k = 1}^K\bfX_k^\top\bfX_k\),
\(\bf A_{(k)} = d\bfI_p - \sum_{c = 1, \dots, K, c \neq k}\bfX_c^\top\bfX_c\),
and
\(\bfX_{-k}^\top\bfy_{-k} = \sum_{c = 1, \dots, K, c \neq k}\bfX_c^\top\bfy_c\).
Then the main computational tasks for fitting a model on the entire
training data and for fitting \(k\) models for cross validation is in
computing \(\bfX_k^\top\bfX_k\) and \(\bfX_{k}^\top\bfy_{k}\) for each
\(k\), which has a total computational complexity of \(O(np^2 + np)\)
for any \(k\). Hence, we can precompute these quantities and the
computation time for the entire cross validation procedure can be
dramatically reduced from the naive procedure of fitting models for each
cross validation fold individually. It is clear that
\(\bfX_k^\top\bfX_k\) and \(\bfX_{k}^\top\bfy_{k}\) can be computed
independently across all \(k\) and hence we can reduce the computational
load even further by computing them in parallel. The techniques
presented here are not applicable to models beyond the linear model,
such as logistic regression.

\section{Extension to logistic
regression}\label{extension-to-logistic-regression}

The logistic regression model is often used when the response of
interest is a binary outcome. The OEM algorithm can be extended to
handle logistic regression models by using a proximal Newton algorithm
similar to that used in the \pkg{glmnet} package and described in
\citet{friedman2010}. OEM can act as a replacement for coordinate
descent in the inner loop in the algorithm described in Section 3 of
\citet{friedman2010}. While we do not present any new algorithmic
results here, for the sake of clarity we will outline the proximal
Newton algorithm of \citet{friedman2010} that we use.

The response variable \(Y\) takes values in \(\{0,1\}\). The logistic
regression model posits the following model for the probability of an
event conditional on predictors: \[
\mu(x) = \mbox{Pr}(Y = 1|x) = \frac{1}{1 + \exp{-(x^\top\coef)}}.
\] Under this model, we compute penalized regression estimates by
maximizing the following penalized log-likelihood with respect to
\(\coef\):

\begin{equation}
\frac{1}{n}\sum_{i = 1}^n \left\{ y_i x_i^\top\coef + \log (1+ \exp(x_i^\top\coef) )   \right\} - P_\lambda(\coef). \label{eqn:pen_lik_logistic}
\end{equation}

Then, given a current estimate \(\hat{\coef}\), we approximate
(\ref{eqn:pen_lik_logistic}) with the following weighted penalized
linear regression:

\begin{equation}
-\frac{1}{2n}\sum_{i = 1}^n w_i\{z_i - x_i^\top\coef \} - P_\lambda(\coef), \label{eqn:approx_pen_lik_logistic}
\end{equation}

where

\begin{align*}
z_i = {} &  x_i^\top\hat{\coef} + \frac{y_i - \hat{\mu}(x_i)}{\hat{\mu}(x_i)(1 - \hat{\mu}(x_i))} \\
w_i = {} & \hat{\mu}(x_i)(1 - \hat{\mu}(x_i))
\end{align*}

and \(\hat{\mu}(x_i)\) is evaluated at \(\hat{\coef}\). Here, \(z_i\)
are the working responses and \(w_i\) are the weights. For each
iteration of the proximal Newton algorithm, we maximize
(\ref{eqn:approx_pen_lik_logistic}) using the OEM algorithm. Similar to
\citet{krishnapuram2005, friedman2010}, we optionally employ an
approximation to the Hessian using an upper-bound of \(w_i = 0.25\) for
all \(i\). This upper bound is often quite efficient for big tall data
settings.

\section[The oem package]{The \pkg{oem} package}\label{the-oem-package}

\subsection[The oem() function]{The \code{oem()} function}\label{the-oem-function}

The function \code{oem()} is the main workhorse of the \pkg{oem}
package.

\begin{CodeChunk}
\begin{CodeInput}
nobs  <- 1e4
nvars <- 25
rho   <- 0.25
sigma <- matrix(rho, ncol = nvars, nrow = nvars)
diag(sigma) <- 1
x     <- mvrnorm(n = nobs, mu = numeric(nvars), Sigma = sigma)
y     <- drop(x %*% c(0.5, 0.5, -0.5, -0.5, rep(0, nvars - 4))) + 
  rnorm(nobs, sd = 3)
\end{CodeInput}
\end{CodeChunk}

The group membership indices for each covariate must be specified for
the group lasso via the argument \code{groups}. The argument
\code{gamma} specifies the \(\gamma\) value for MCP. The function
\code{plot.oemfit()} allows the user to plot the estimated coefficient
paths. Its argument \code{which.model} allows the user to select model
to be plotted.

\begin{CodeChunk}
\begin{CodeInput}
fit   <- oem(x = x, y = y, penalty = c("lasso", "mcp", "grp.lasso"),
  gamma = 2, groups = rep(1:5, each = 5),
  lambda.min.ratio = 1e-3)
\end{CodeInput}
\end{CodeChunk}

\begin{CodeChunk}
\begin{CodeInput}
par(mar=c(5, 5, 5, 3) + 0.1)
layout(matrix(1:3, ncol = 3))
plot(fit, which.model = 1, xvar = "lambda", 
  cex.main = 3, cex.axis = 1.25, cex.lab = 2)
plot(fit, which.model = 2, xvar = "lambda",
  cex.main = 3, cex.axis = 1.25, cex.lab = 2)
plot(fit, which.model = 3, xvar = "lambda",
  cex.main = 3, cex.axis = 1.25, cex.lab = 2)
\end{CodeInput}
\begin{figure}[H]

{\centering \includegraphics{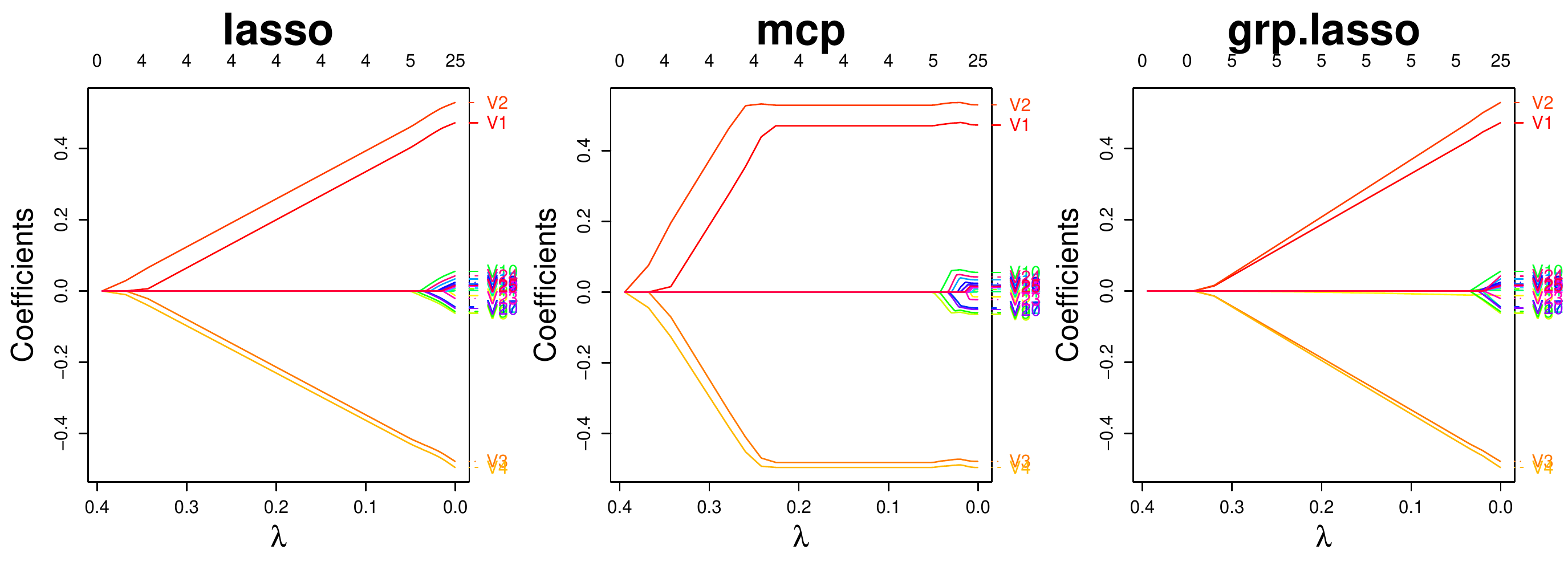} 

}

\caption[The plots above depict estimated coefficient paths for the lasso, MCP, and group lasso]{The plots above depict estimated coefficient paths for the lasso, MCP, and group lasso.}\label{fig:plot_path}
\end{figure}
\end{CodeChunk}

To compute the loss function in addition to the estimated coefficients,
the argument \code{compute.loss} must be set to \code{TRUE} like the
following:

\begin{CodeChunk}
\begin{CodeInput}
fit   <- oem(x = x, y = y, penalty = c("lasso", "mcp", "grp.lasso"),
  gamma = 2, groups = rep(1:5, each = 5),
  lambda.min.ratio = 1e-3, compute.loss = TRUE)
\end{CodeInput}
\end{CodeChunk}

By default, \code{compute.loss} is set to \code{FALSE} because it adds a
large computational burden, especially when many penalties are input.
The function \code{logLik.oemfit()} can be used in complement with
fitted with \code{oem()} objects with \code{compute.loss = TRUE} with
the model specified using the \code{which.model} argument like the
following:

\begin{CodeChunk}
\begin{CodeInput}
logLik(fit, which.model = 2)[c(1, 25, 50, 100)]
\end{CodeInput}
\begin{CodeOutput}
[1] -14189.39 -13804.72 -13795.76 -13795.11
\end{CodeOutput}
\end{CodeChunk}

\subsection[Fitting multiple
penalties]{Fitting multiple
penalties}\label{fitting-multiple-penalties}

The OEM algorithm is well-suited to quickly estimate a solution path for
multiple penalties simultaneously for the linear model if the number of
variables is not too large, often when the number of variables is
several thousand or fewer, provided the number of observations is larger
than the number of variables. Ideally the number of observations should
be at least ten times larger than the number of variables for best
performance. Once the quantities \(\bf A\) and \(\bfX^\top\bfy\) are
computed initially, then the remaining computational complexity for OEM
for a given model and tuning parameter is just \(O(p^2)\) per iteration.
To demonstrate the efficiency, consider the following simulated example:

\begin{CodeChunk}
\begin{CodeInput}
nobs   <- 1e6
nvars  <- 100
rho    <- 0.25
sigma  <- matrix(rho, ncol = nvars, nrow = nvars)
diag(sigma) <- 1
x2     <- mvrnorm(n = nobs, mu = numeric(nvars), Sigma = sigma)
y2     <- drop(x2 %*% c(0.5, 0.5, -0.5, -0.5, rep(0, nvars - 4))) + 
  rnorm(nobs, sd = 5)
\end{CodeInput}
\end{CodeChunk}

\begin{CodeChunk}
\begin{CodeInput}
mb <- microbenchmark(
         "oem[lasso]" = oem(x = x2, y = y2, 
                            penalty = c("lasso"),
                            gamma = 3,
                            groups = rep(1:20, each = 5)),
         "oem[all]"   = oem(x = x2, y = y2, 
                            penalty = c("lasso", "mcp", 
                                        "grp.lasso", "scad"),
                            gamma = 3,
                            groups = rep(1:20, each = 5)),
         times = 10L)
print(mb, digits = 3)
\end{CodeInput}
\begin{CodeOutput}
Unit: seconds
       expr  min   lq mean median   uq  max neval cld
 oem[lasso] 2.38 2.39 2.41   2.42 2.44 2.46    10  a 
   oem[all] 2.88 2.89 2.92   2.91 2.93 3.00    10   b
\end{CodeOutput}
\end{CodeChunk}

\subsection[Parallel support via
OpenMP]{Parallel support via
\pkg{OpenMP}}\label{parallel-support-via-openmp}

As noted in Section \ref{parallelization-and-fast-cross-validation}, the
key quantities necessary for the OEM algorithm can be computed in
parallel. By specifying \texttt{ncores} to be a value greater than 1,
the \code{oem()} function automatically employs \pkg{OpenMP}
\citep{openmp15} to compute \(\bf A\) and \(\bfX^\top\bfy\) in parallel.
Due to memory access inefficiencies in breaking up the computation of
\(\bf A\) into pieces, using multiple cores does not speed up
computation linearly. It is typical for \pkg{OpenMP} not to result in
linear speedups, especially on Windows machines, due to its overhead
costs. Furthermore, if a user does not have OpenMP on their machine, the
\pkg{oem} package will still run normally on one core. In the following
example, we can see a slight benefit from invoking the use of extra
cores.

\begin{CodeChunk}
\begin{CodeInput}
nobs   <- 1e5
nvars  <- 500
rho    <- 0.25
sigma  <- rho ** abs(outer(1:nvars, 1:nvars, FUN = "-"))
x2     <- mvrnorm(n = nobs, mu = numeric(nvars), Sigma = sigma)
y2     <- drop(x2 %*% c(0.5, 0.5, -0.5, -0.5, rep(0, nvars - 4))) + 
  rnorm(nobs, sd = 5)
mb <- microbenchmark(
         "oem"           = oem(x = x2, y = y2, 
                               penalty = c("lasso", "mcp", 
                                           "grp.lasso", "scad"),
                               gamma = 3,
                               groups = rep(1:20, each = 25)),
         "oem[parallel]" = oem(x = x2, y = y2, 
                               ncores = 2,
                               penalty = c("lasso", "mcp", 
                                           "grp.lasso", "scad"),
                               gamma = 3,
                               groups = rep(1:20, each = 25)),
         times = 10L)
print(mb, digits = 3)
\end{CodeInput}
\begin{CodeOutput}
Unit: seconds
          expr  min   lq mean median   uq  max neval cld
           oem 4.80 4.85 4.93   4.93 5.00 5.07    10   b
 oem[parallel] 3.72 3.84 3.98   3.95 4.11 4.31    10  a 
\end{CodeOutput}
\end{CodeChunk}

\subsection[The cv.oem() function]{The \code{cv.oem()} function}\label{the-cv.oem-function}

The \code{cv.oem()} function is used for cross validation of penalized
models fitted by the \code{oem()} function. It does not use the method
described in Section \ref{parallelization-and-fast-cross-validation} and
hence can be used for models beyond the linear model. It can also
benefit from parallelization using either \pkg{OpenMP} or the
\pkg{foreach} package. For the former, one only need specify the
\code{ncores} argument of \code{cv.oem()} and computation of the key
quantities for OEM are computed in parallel for each cross validation
fold. With the \pkg{foreach} package \citep{foreach}, cores must be
``registered'' in advance using \pkg{doParallel} \citep{doParallel},
\pkg{doMC} \citep{doMC}, or otherwise. Each cross validation fold is
computed on a separate core, which may be more efficient depending on
the user's hardware.

\begin{CodeChunk}
\begin{CodeInput}
cvfit <- cv.oem(x = x, y = y, 
  penalty = c("lasso", "mcp", "grp.lasso"),
  gamma  = 2, groups = rep(1:5, each = 5), 
  nfolds = 10)
\end{CodeInput}
\end{CodeChunk}

The best performing model and its corresponding best tuning parameter
can be accessed via:

\begin{CodeChunk}
\begin{CodeInput}
cvfit$best.model
\end{CodeInput}
\begin{CodeOutput}
[1] "mcp"
\end{CodeOutput}
\begin{CodeInput}
cvfit$lambda.min
\end{CodeInput}
\begin{CodeOutput}
[1] 0.0739055
\end{CodeOutput}
\end{CodeChunk}

A summary method is available as \code{summary.cv.oem()}, similar to the
summary function of \code{cv.ncvreg()} of the \pkg{ncvreg} package,
which prints output from all of the cross validated models. It can be
used like the following:

\begin{CodeChunk}
\begin{CodeInput}
summary(cvfit)
\end{CodeInput}
\begin{CodeOutput}
lasso-penalized linear regression with n=10000, p=25
At minimum cross-validation error (lambda=0.0242):
-------------------------------------------------
  Nonzero coefficients: 13
  Cross-validation error (Mean-Squared Error): 9.11
  Scale estimate (sigma): 3.018

<===============================================>

mcp-penalized linear regression with n=10000, p=25
At minimum cross-validation error (lambda=0.0739):
-------------------------------------------------
  Nonzero coefficients: 5
  Cross-validation error (Mean-Squared Error): 9.10
  Scale estimate (sigma): 3.016

<===============================================>

grp.lasso-penalized linear regression with n=10000, p=25
At minimum cross-validation error (lambda=0.0242):
-------------------------------------------------
  Nonzero coefficients: 16
  Cross-validation error (Mean-Squared Error): 9.10
  Scale estimate (sigma): 3.017
\end{CodeOutput}
\end{CodeChunk}

Predictions from any model (\code{which.model = 2}) or the best of all
models (\code{which.model = "best.model"}) using \code{cv.oem} objects.
The tuning parameter is specified via the argument \code{s}, which can
take numeric values or \code{"lambda.min"} for the best tuning parameter
or \code{"lambda.1se"} for a good but more conservative tuning
parameter.

\begin{CodeChunk}
\begin{CodeInput}
predict(cvfit, newx = x[1:3,], which.model = "best.model", s = "lambda.min")
\end{CodeInput}
\begin{CodeOutput}
           [,1]
[1,] -0.2233264
[2,]  0.2211386
[3,]  0.4760399
\end{CodeOutput}
\end{CodeChunk}

\subsection[The xval.oem() function]{The \code{xval.oem()} function}\label{the-xval.oem-function}

The \code{xval.oem()} function is much like \code{cv.oem()} but is
limited to use for linear models only. It is significantly faster than
\code{cv.oem()}, as it uses the method described in Section
\ref{parallelization-and-fast-cross-validation}. Whereas \code{cv.oem()}
functions by repeated calls to \code{oem()}, all of the primary
computation in \code{xval.oem()} is carried out in \proglang{C++}. We
chose to keep the \code{xval.oem()} function separate from the
\code{cv.oem()} because the underlying code between the two methods is
vastly different and furthermore because \code{xval.oem()} is not
available for logistic regression models.

\begin{CodeChunk}
\begin{CodeInput}
xvalfit <- xval.oem(x = x, y = y, 
  penalty = c("lasso", "mcp", "grp.lasso"),
  gamma  = 2, groups = rep(1:5, each = 5), 
  nfolds = 10)

yrng <- range(c(unlist(xvalfit$cvup), unlist(xvalfit$cvlo)))
layout(matrix(1:3, ncol = 3))
par(mar=c(5, 5, 5, 3) + 0.1)
plot(xvalfit, which.model = 1, ylim = yrng, 
  cex.main = 3, cex.axis = 1.25, cex.lab = 2)
plot(xvalfit, which.model = 2, ylim = yrng,
  cex.main = 3, cex.axis = 1.25, cex.lab = 2)
plot(xvalfit, which.model = 3, ylim = yrng, 
  cex.main = 3, cex.axis = 1.25, cex.lab = 2)
\end{CodeInput}
\begin{figure}

{\centering \includegraphics{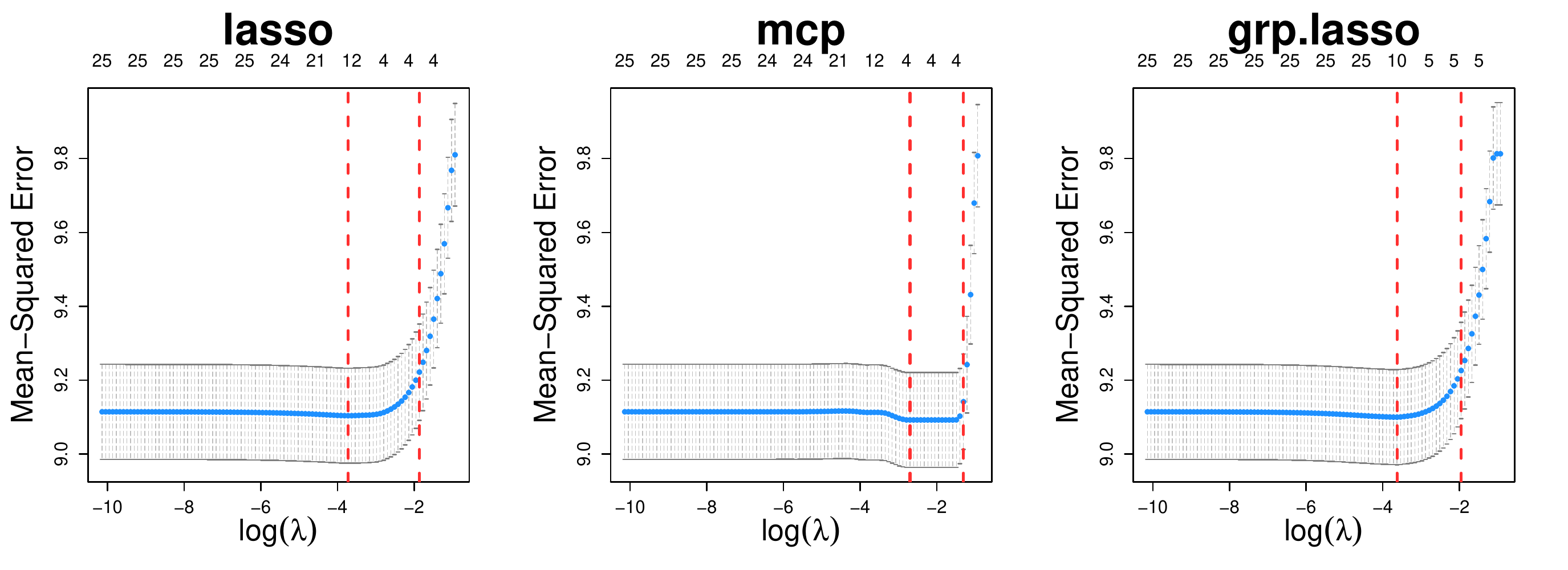} 

}

\caption[Depicted above are the cross validated mean squared prediction errors for paths of tuning parameters for the lasso, MCP, and group lasso]{Depicted above are the cross validated mean squared prediction errors for paths of tuning parameters for the lasso, MCP, and group lasso.}\label{fig:xvalone}
\end{figure}
\end{CodeChunk}

\subsection[OEM with precomputation for linear models with the oem.xtx()
function]{OEM with precomputation for linear models with the \code{oem.xtx()}
function}\label{oem-with-precomputation-for-linear-models-with-the-oem.xtx-function}

The key quantities, \(\bfX^\top\bfX\) and \(\bfX^\top\bfy\) can be
computed in parallel, and when data are stored across a large cluster,
their computation can be performed in a straightforward manner. When
they are available, the \code{oem.xtx()} function can be used to carry
out the OEM algorithm based on these quantities instead of on the full
design matrix \(\bfX\) and the full response vector \(\bfy\). All
methods available to objects fitted by \code{oem()} are also available
to objects fitted by \code{oem.xtx()}.

\begin{CodeChunk}
\begin{CodeInput}
xtx <- crossprod(x) / nrow(x)
xty <- crossprod(x, y) / nrow(x)
\end{CodeInput}
\end{CodeChunk}

\begin{CodeChunk}
\begin{CodeInput}
fitxtx <- oem.xtx(xtx, xty, 
  penalty = c("lasso", "mcp", "grp.lasso"),
  gamma = 2, groups = rep(1:5, each = 5))
\end{CodeInput}
\end{CodeChunk}

\subsection[Out-of-Memory computation with the big.oem()
function]{Out-of-Memory computation with the \code{big.oem()}
function}\label{out-of-memory-computation-with-the-big.oem-function}

Standard \proglang{R} objects are stored in memory and thus, when a
design matrix is too large for memory it cannot be used for computation
in the standard way. The \pkg{bigmemory} package offers objects which
point to data stored on disk and thus allows users to bypass memory
limitations. It also provides access to \proglang{C++} objects which do
the same. These objects are highly efficient due to memory mapping,
which is a method of mapping a data file to virtual memory and allows
for efficient moving of data in and out of memory from disk. For further
details on memory mapping, we refer readers to
\citet{bovet2005, kane2013}. The \code{big.oem()} function allows for
out-of-memory computation by linking \pkg{Eigen} matrix objects to data
stored on disk via \pkg{bigmemory}.

The standard approach for loading the data from \code{SEXP} objects
(\code{X_} in the example below) to \pkg{Eigen} matrix objects (\code{X}
in the example below) in \proglang{C++} looks like:

\begin{CodeChunk}
\begin{CodeInput}
using Eigen::Map;
using Eigen::MatrixXd;
const Map<MatrixXd> X(as<Map<MatrixXd> >(X_));
\end{CodeInput}
\end{CodeChunk}

To instead map from an object which is a pointer to data on disk, we
first need to load the \pkg{bigmemory} headers:

\begin{CodeChunk}
\begin{CodeInput}
#include <bigmemory/MatrixAccessor.hpp>
#include <bigmemory/BigMatrix.h>
\end{CodeInput}
\end{CodeChunk}

Then we link the pointer (passed from \proglang{R} to \proglang{C++} as
\code{X_} and set as \code{bigPtr} below) to data to an \pkg{Eigen}
matrix object via:

\begin{CodeChunk}
\begin{CodeInput}
XPtr<BigMatrix> bigPtr(X_);
const Map<MatrixXd>  X = Map<MatrixXd>
    ((double *)bigPtr->matrix(), bigPtr->nrow(), bigPtr->ncol()  );
\end{CodeInput}
\end{CodeChunk}

The remaining computation for OEM is carried out similarly as for
\code{oem()}, yet here the object \code{X} is not stored in memory.

To test out \code{big.oem()} and demonstrate its memory usage profile we
simulate a large dataset and save it as a ``filebacked''
\code{big.matrix} object from the \pkg{bigmemory} package.

\begin{CodeChunk}
\begin{CodeInput}
nobs     <- 1e6
nvars    <- 250
bkFile   <- "big_matrix.bk"
descFile <- "big_matrix.desc"
big_mat  <- filebacked.big.matrix(nrow           = nobs, 
                                  ncol           = nvars, 
                                  type           = "double",  
                                  backingfile    = bkFile, 
                                  backingpath    = ".", 
                                  descriptorfile = descFile,
                                  dimnames       = c(NULL, NULL))

for (i in 1:nvars) 
{
    big_mat[, i] = rnorm(nobs)
}

yb <- rnorm(nobs, sd = 5)
\end{CodeInput}
\end{CodeChunk}

Using the \code{profvis()} function of the \pkg{profvis} package
\citep{profvis}, we can see that no copies of the design matrix are made
at any point. Furthermore, a maximum of 173 Megabytes are used by the
\proglang{R} session during this simulation, whereas the size of the
design matrix is 1.9 Gigabytes. The following code generates an
interactive \proglang{html} visualization of the memory usage of
\code{big.oem()} line-by-line:

\begin{CodeChunk}
\begin{CodeInput}
profvis::profvis({
    bigfit <- big.oem(x       = big_mat, y = yb, 
                      penalty = c("lasso", "grp.lasso", "mcp", "scad"),
                      gamma   = 3,
                      groups  = rep(1:50, each = 5))
})
\end{CodeInput}
\end{CodeChunk}

Here we save a copy of the design matrix in memory for use by
\code{oem()}:

\begin{CodeChunk}
\begin{CodeInput}
xb <- big_mat[,]

print(object.size(xb), units = "Mb")
\end{CodeInput}
\begin{CodeOutput}
1907.3 Mb
\end{CodeOutput}
\begin{CodeInput}
print(object.size(big_mat), units = "Mb")
\end{CodeInput}
\begin{CodeOutput}
0 Mb
\end{CodeOutput}
\end{CodeChunk}

The following benchmark for on-disk computation is on a system with a
hard drive with 7200 RPM, 16MB Cache, and SATA 3.0 Gigabytes per second
(a quite modest setup compared with a system with a solid state drive).
Even without a solid state drive we pay little time penalty for
computing on disk over computing in memory.

\begin{CodeChunk}
\begin{CodeInput}
mb <- microbenchmark(
         "big.oem" = big.oem(x       = big_mat, y = yb, 
                             penalty = c("lasso", "grp.lasso", 
                                         "mcp", "scad"),
                             gamma   = 3,
                             groups  = rep(1:50, each = 5)),
         "oem"     =     oem(x       = xb, y = yb, 
                             penalty = c("lasso", "grp.lasso", 
                                         "mcp", "scad"),
                             gamma   = 3,
                             groups  = rep(1:50, each = 5)),
         times = 10L)

print(mb, digits = 3)
\end{CodeInput}
\begin{CodeOutput}
Unit: seconds
    expr  min   lq  mean median    uq   max neval cld
 big.oem 8.46 8.51  8.73   8.57  8.69  9.73    10  a 
     oem 9.81 9.85 10.36  10.08 10.65 12.03    10   b
\end{CodeOutput}
\end{CodeChunk}

\subsection[Sparse matrix support]{Sparse matrix support}\label{sparse-matrix-support}

The \code{oem()} and \code{cv.oem()} functions can accept sparse design
matrices as provided by the \code{CsparseMatrix} class of objects of the
\pkg{Matrix} package \citep{Matrix}. If the design matrix provided has a
high degree of sparsity, using a \code{CsparseMatrix} object can result
in a substantial computational speedup and reduction in memory usage.

\begin{CodeChunk}
\begin{CodeInput}
library(Matrix)
n.obs  <- 1e5
n.vars <- 200
true.beta <- c(runif(15, -0.25, 0.25), rep(0, n.vars - 15))
xs <- rsparsematrix(n.obs, n.vars, density = 0.01)
ys <- rnorm(n.obs, sd = 3) + as.vector(xs %*% true.beta)
x.dense <- as.matrix(xs)

mb <- microbenchmark(fit   = oem(x = x.dense, y = ys, 
                             penalty = c("lasso", "grp.lasso"), 
                             groups = rep(1:40, each = 5)),
                     fit.s = oem(x = xs, y = ys, 
                                 penalty = c("lasso", "grp.lasso"), 
                                 groups = rep(1:40, each = 5)),
                     times = 10L)

print(mb, digits = 3)
\end{CodeInput}
\begin{CodeOutput}
Unit: milliseconds
  expr   min    lq  mean median    uq   max neval cld
   fit 669.9 672.5 679.3  680.3 682.2 690.6    10   b
 fit.s  63.2  64.1  65.5   65.3  66.8  68.2    10  a 
\end{CodeOutput}
\end{CodeChunk}

\subsection[API comparison with
glmnet]{API comparison with
\pkg{glmnet}}\label{api-comparison-with-glmnet}

The application program interface (API) of the \pkg{oem} package was
designed to be familiar to users of the \pkg{glmnet} package. Data ready
for use by \code{glmnet()} can be used directly by \code{oem()}. Most of
the arguments are the same, except the \code{penalty} argument and other
arguments relevant to the various penalties available in \code{oem()}.

Here we fit linear models with a lasso penalty using \code{oem()} and
\code{glmnet()}:

\begin{CodeChunk}
\begin{CodeInput}
oem.fit    <- oem(x = x, y = y, penalty = "lasso")
glmnet.fit <- glmnet(x = x, y = y)
\end{CodeInput}
\end{CodeChunk}

Here we fit linear models with a lasso penalty using \code{oem()} and
\code{glmnet()} with sparse design matrices:

\begin{CodeChunk}
\begin{CodeInput}
oem.fit.sp    <- oem(x = xs, y = ys, penalty = "lasso")
glmnet.fit.sp <- glmnet(x = xs, y = ys)
\end{CodeInput}
\end{CodeChunk}

Now we make predictions using the fitted model objects from both
packages:

\begin{CodeChunk}
\begin{CodeInput}
preds.oem    <- predict(oem.fit, newx = x)
preds.glmnet <- predict(glmnet.fit, newx = x)
\end{CodeInput}
\end{CodeChunk}

We now plot the coefficient paths using both fitted model objects:

\begin{CodeChunk}
\begin{CodeInput}
plot(oem.fit, xvar = "norm")
plot(glmnet.fit, xvar = "norm")
\end{CodeInput}
\end{CodeChunk}

We now fit linear models with a lasso penalty and select the tuning
parameter with cross validation using \code{cv.oem()},
\code{xval.oem()}, and \code{cv.glmnet()}:

\begin{CodeChunk}
\begin{CodeInput}
oem.cv.fit    <- cv.oem(x = x, y = y, penalty = "lasso")
oem.xv.fit    <- xval.oem(x = x, y = y, penalty = "lasso")
glmnet.cv.fit <- cv.glmnet(x = x, y = y)
\end{CodeInput}
\end{CodeChunk}

We now plot the cross validation errors using all fitted cross
validation model objects:

\begin{CodeChunk}
\begin{CodeInput}
plot(oem.cv.fit)
plot(oem.xv.fit)
plot(glmnet.cv.fit)
\end{CodeInput}
\end{CodeChunk}

We now make predictions using the best tuning parameter according to
cross validation error using all fitted cross validation model objects:

\begin{CodeChunk}
\begin{CodeInput}
preds.cv.oem    <- predict(oem.cv.fit, newx = x, s = "lambda.min")
preds.xv.oem    <- predict(oem.xv.fit, newx = x, s = "lambda.min")
preds.cv.glmnet <- predict(glmnet.cv.fit, newx = x, s = "lambda.min")
\end{CodeInput}
\end{CodeChunk}

\section[Timings]{Timings}\label{timings}

Extensive numerical studies are conducted in \citet{xiong16} regarding
the OEM algorithm for computation time for paths of tuning parameters
for various penalties, so in the following simulation studies, we will
focus on computation time for the special features of the \pkg{oem}
package such as cross validation and sparse matrix support. All
simulations are run on a 64-bit machine with an Intel Xeon E5-2470 2.30
GHz CPU and 128 Gigabytes of main memory and a Linux operating system.

\subsection[Cross validation]{Cross validation}\label{cross-validation}

In this section we will compare the computation time of the
\code{cv.oem()} and \code{xval.oem()} for various penalties, both
individually and simultanously, with the cross validation functions from
other various packages, including \pkg{glmnet}, \pkg{ncvreg},
\pkg{grpreg}, \pkg{gglasso}, and the \proglang{Python} package
\pkg{sklearn} \citep{pedregosa2011}. The model class we use from
\pkg{sklearn} is \code{LassoCV}, which performs cross validation for
lasso linear models. In particular, we focus on the comparison with
\pkg{glmnet}, as it has been carefully developed with computation time
in mind and has long been the gold standard for computational
performance.

In the simulation setup, we generate the design matrix from a
multivariate normal distribution with covariance matrix
\((\sigma_{ij}) = 0.5 ^ {|i - j|}\). Responses are generated from the
following model: \[
\bfy = \bfX\coef + \boldsymbol\epsilon
\] where the first five elements of \(\coef\) are
\((-0.5, -0.5, 0.5, 0.5, 1)\) and the remaining are zero and
\(\boldsymbol\epsilon\) is an independent mean zero normal random
variable with standard deviation 2. The total number of observations
\(n\) is set to \(10^5\) and \(10^6\), the number of variables \(p\) is
varied from 50 to 500, and the number of folds for cross validation is
set to 10. For grouped regularization, groups of variables of size 25
are chosen contiguously. For the MCP regularization, the tuning
parameter \(\gamma\) is chosen to be 3. Each method is fit using the
same sequence of 100 values of the tuning parameter \(\lambda\) and
convergence is specified to be the same level of precision for each
method. The computation times for the \code{glmnet} and \code{oem}
functions without cross validation are given as a reference point.

From the results in Figure \ref{fig:plotCV}, both the \code{xval.oem()}
and \code{cv.oem()} functions are competitive with all other cross
validation alternatives. Surprisingly, the \code{xval.oem()} function
for the lasso penalty only is competitive with and in many scenarios is
even faster than \code{glmnet}, which does not perform cross validation.
The \code{xval.oem()} is clearly faster than \code{cv.oem()} in all
scenarios. In many scenarios \code{cv.oem()} takes at least 6 times
longer than \code{xval.oem()}. Both cross validation functions from the
\pkg{oem} package are nearly as fast in computing for three penalties
simultaneously as they are for just one. We have found in general that
for any scenario where \(n >> p\) and cross validation is required, it
is worth considering \pkg{oem} as a fast alternative to other packages.
In particular, a rough rule of thumb is that \pkg{oem} is advantageous
when \(n > 10p\), however, \pkg{oem} may be advantageous with fewer
observations than this for penalties other than the lasso, such as MCP,
SCAD, and group lasso.

\begin{CodeChunk}
\begin{figure}

{\centering \includegraphics{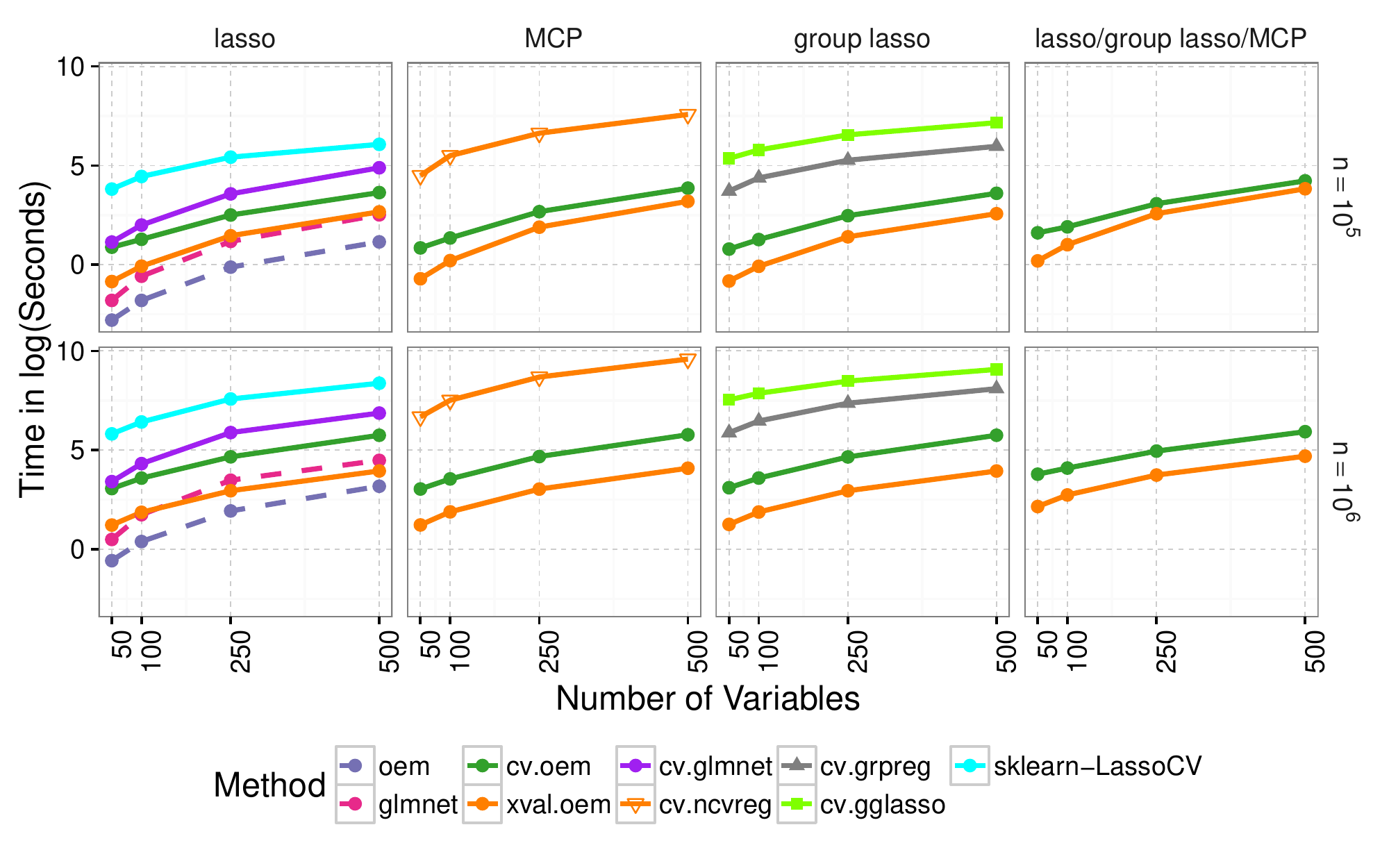} 

}

\caption[Depicted above is the average computing time over ten runs in log(Seconds) for the functions that perform cross validation for selection of the tuning parameter for various penalized regression methods]{Depicted above is the average computing time over ten runs in log(Seconds) for the functions that perform cross validation for selection of the tuning parameter for various penalized regression methods.}\label{fig:plotCV}
\end{figure}
\end{CodeChunk}

Now we test the impact of parallelization on the computation time of
\code{xval.oem()} using the same simulation setup as above except now we
additionally vary the number of cores used. From the results in Figure
\ref{fig:plotCVPar}, we can see that using parallelization through
\pkg{OpenMP} helps to some degree. However, it is important to note that
using more cores than the number of cross validation folds will unlikely
result in better computation time than the same number of cores as the
number of folds due to how parallelization is implemented.

\begin{CodeChunk}
\begin{figure}

{\centering \includegraphics{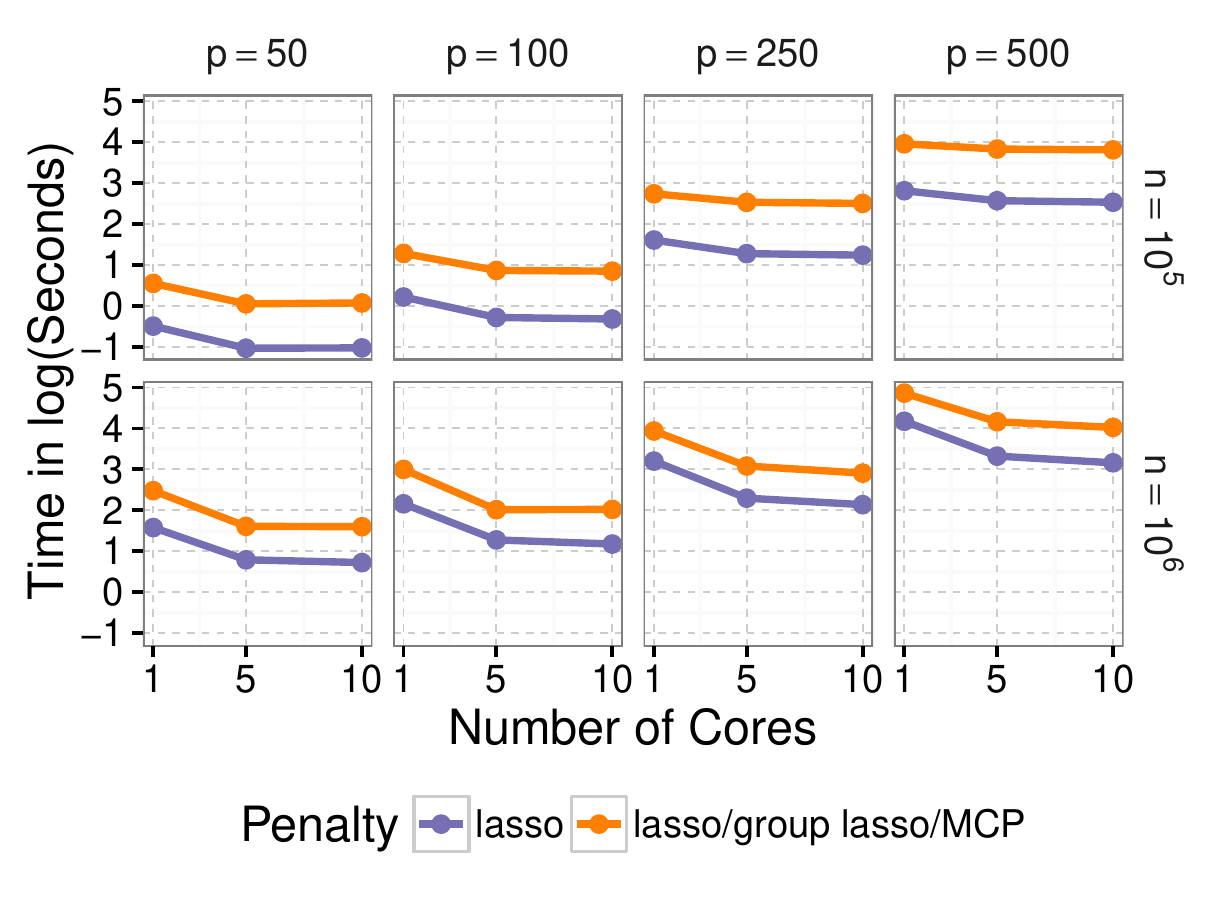} 

}

\caption{Depicted above is the average computing time over ten runs in log(Seconds) for the \code{xval.oem()} function with a varying number of cores. Results are shown for just one penalty (lasso) and three penalties fit simultaneously (lasso, group lasso, and MCP) with 100 values for the tuning parameter $\lambda$ for each penalty.}\label{fig:plotCVPar}
\end{figure}
\end{CodeChunk}

\subsection[Sparse matrices]{Sparse matrices}\label{sparse-matrices}

In the following simulated example, sparse design matrices are generated
using the function \code{rsparsematrix()} of the \pkg{Matrix} package
with nonzero entries generated from an independent standard normal
distribution. The proportion of zero entries in the design matrix is set
to 0.99, 0.95, and 0.9. The total number of observations \(n\) is set to
\(10^5\) and \(10^6\) and the number of variables \(p\) is varied from
250 to 1000. Responses are generated from the same model as in Section
\ref{cross-validation}. Each method is fit using the same sequence of
100 values of the tuning parameter \(\lambda\) and convergence is
specified to be the same level of precision for each method. The authors
are not aware of penalized regression packages other than \pkg{glmnet}
that provide support for sparse matrices, so we just compare \pkg{oem}
with \pkg{glmnet}. From the results in Figure \ref{fig:plotSPARSE}, it
can be seen that \code{oem()} is superior in computation time for very
tall data settings with a high degree of sparsity of the design matrix.

\begin{CodeChunk}
\begin{figure}

{\centering \includegraphics{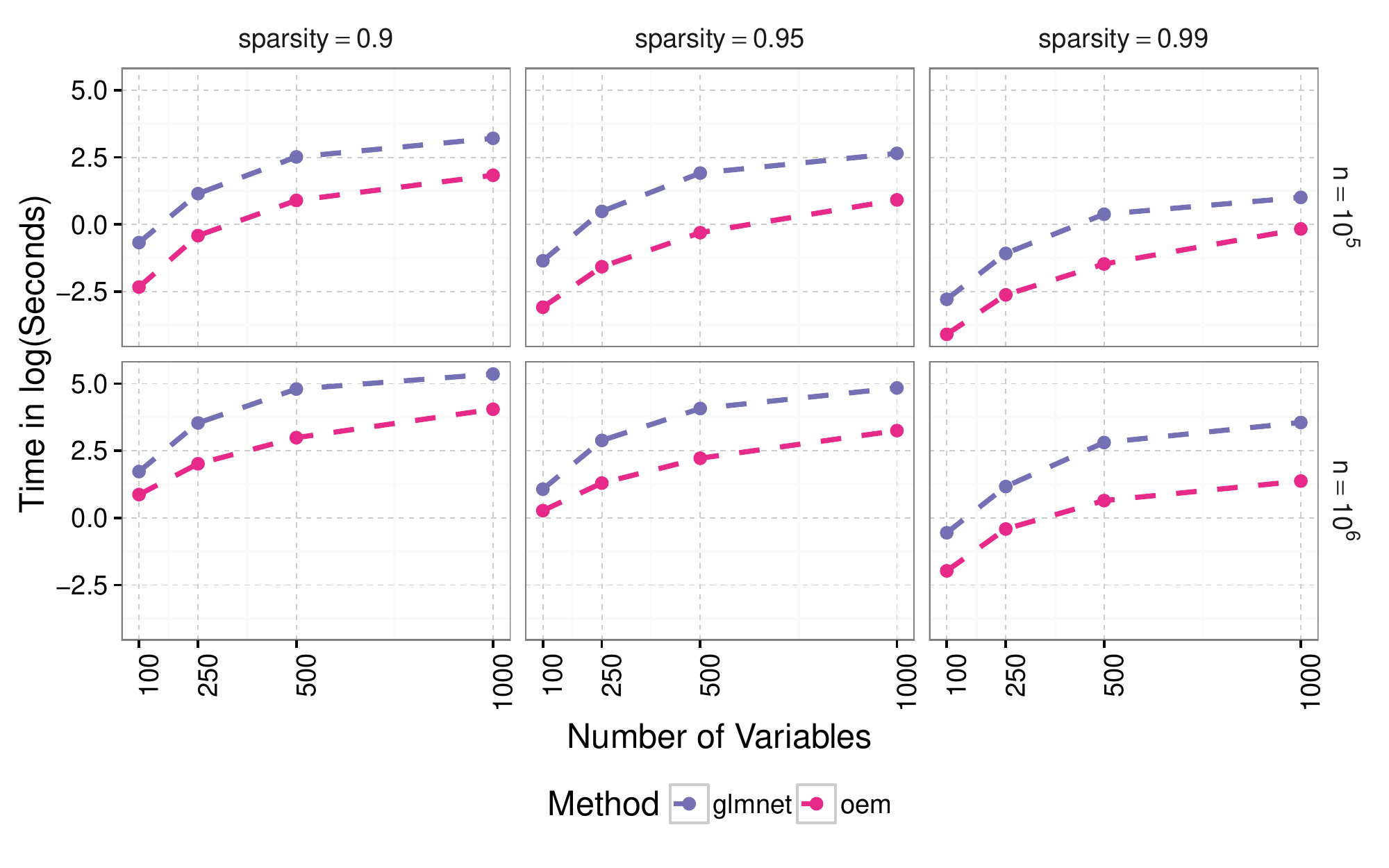} 

}

\caption[Depicted above is the average computing time over ten runs in log(Seconds) for computation of lasso linear regression models using sparse matrices]{Depicted above is the average computing time over ten runs in log(Seconds) for computation of lasso linear regression models using sparse matrices.}\label{fig:plotSPARSE}
\end{figure}
\end{CodeChunk}

\subsection[Penalized logistic
regression]{Penalized logistic
regression}\label{penalized-logistic-regression}

In this simulation, we generate the design matrix from a multivariate
normal distribution with covariance matrix
\((\sigma_{ij}) = 0.5 ^ {|i - j|}\). Responses are generated from the
following model: \[
\mbox{Pr}(Y_i = \boldsymbol 1|x_i) = \frac{1}{1 + \exp{(-x_i^\top\coef)}}, \mbox{ for } i = 1, \dots, n
\] where the first five elements of \(\coef\) are
\((-0.5, -0.5, 0.5, 0.5, 1)\) and the remaining are zero. The total
number of observations \(n\) is set to \(10^4\) and \(10^5\), the number
of variables \(p\) is varied from 50 to 500. For grouped regularization,
groups of variables of size 25 are chosen contiguously. For the MCP
regularization, the tuning parameter \(\gamma\) is chosen to be 3. Each
method is fit using the same sequence of 25 values of the tuning
parameter \(\lambda\). Both \code{glmnet()} and \code{oem()} allow for a
Hessian upper-bound approximation for logistic regression models, so in
this simulation we compare both the vanilla versions of each in addition
to the Hessian upper-bound versions. For logistic regression models,
unlike linear regression, the various functions compared use vastly
different convergence criterion, so care was taken to set the
convergence thresholds for the different methods to achieve a similar
level of precision.

The average computing times over 10 runs are displayed in Figure
\ref{fig:plotBINOMIAL}. The Hessian upper-bound versions of
\code{glmnet()} and \code{oem()} are given by the dashed lines. The
\code{glmnet()} function is the fastest for lasso-penalized logistic
regression, but \code{oem()} with an without the use of a Hessian
upper-bound is nearly as fast as \code{glmnet()} when the sample size
gets larger. For the MCP and group lasso, \code{oem()} with the Hessian
upper-bound option is the fastest across the vast majority of simulation
settings. In general, the Hessian upper-bound is most advantageous when
the number of variables is moderately large.

\begin{CodeChunk}
\begin{figure}

{\centering \includegraphics{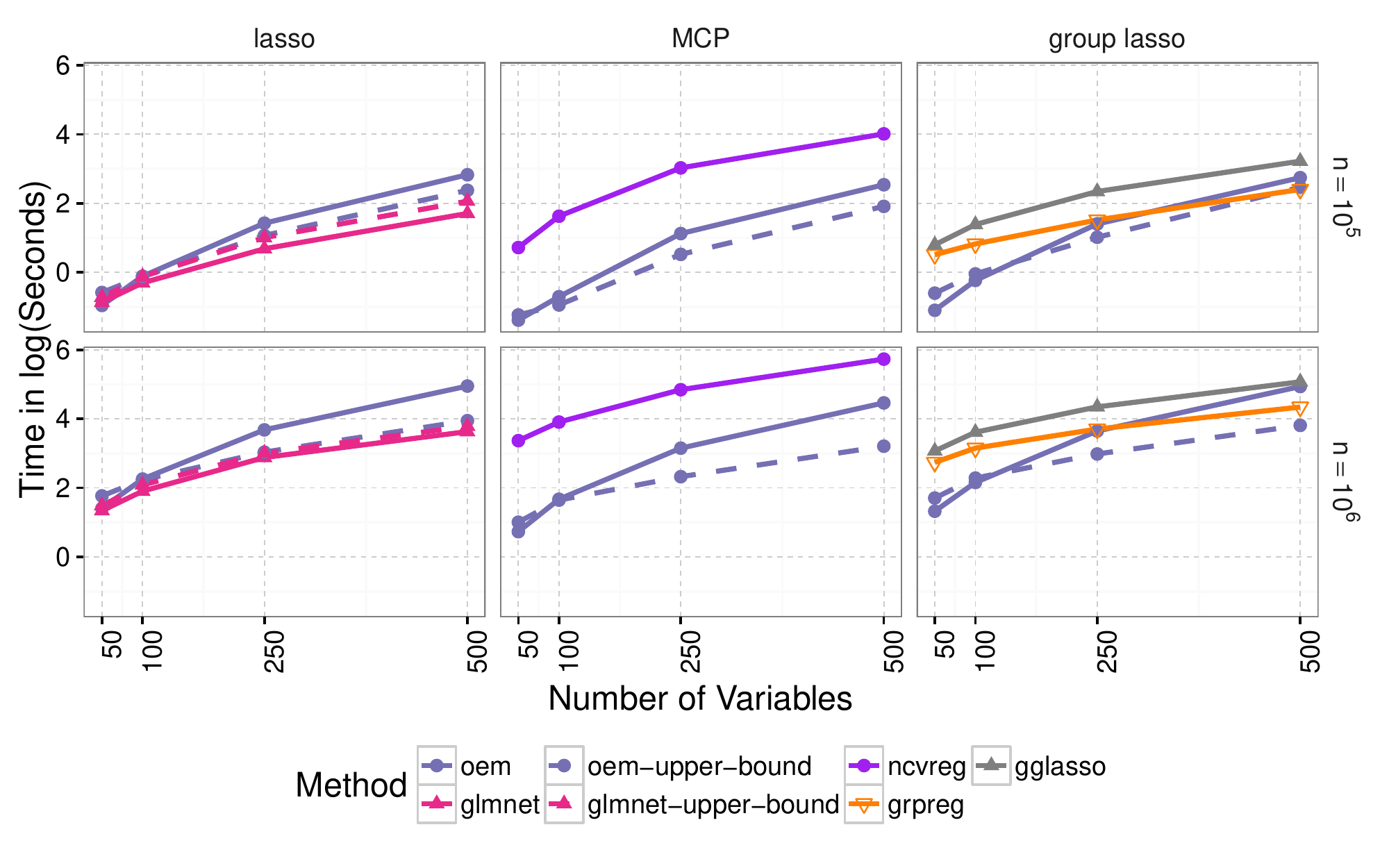} 

}

\caption[Depicted above is the average computing time over ten runs in log(Seconds) for the functions that penalized estimates for binomial regression models]{Depicted above is the average computing time over ten runs in log(Seconds) for the functions that penalized estimates for binomial regression models.}\label{fig:plotBINOMIAL}
\end{figure}
\end{CodeChunk}

\subsection[Quality of solutions]{Quality of solutions}\label{quality-of-solutions}

In this section we present information about the quality of solutions
provided by each of the compared packages. Using the same simulation
setup and precisions for convergence as the cross validation and
binomial simulations, we investigate the numerical precision of the
methods by presenting the objective function values corresponding to the
given solutions. Specifically, we look at the difference in the
objective function values between \code{oem()} and the comparative
functions. A Negative value here indicates that the \code{oem()}
function results in solutions with a smaller value of the objective
function. Since each simulation is run on a sequence of tuning parameter
values we present the differences in objective functions averaged over
the tuning parameters. Since the objective function values across the
different tuning parameter values are on the same scale, this comparison
is reasonable. While the \code{grpreg()} function from the \pkg{grpreg}
package provides solutions for the group lasso, it minimizes a slightly
modified objective function, wherein the covariates are orthonormalized
within groups as described in Section 2.1 of \citet{breheny2015}. Due to
this fact we do not compare the objective function values assocated with
results returned by \code{grpreg()}.

The objective function value differences for the linear model
simulations are presented in Table \ref{table:loss_linear}. We can see
that the solutions provided by \code{oem()} are generally more precise
than those of other functions except \code{ncvreg()} for the MCP.
However, the differences between the solutions provided by \code{oem()}
and \code{ncvreg()} for the MCP are close to machine precision. The
objective function value differences for the logistic model simulations
are presented in Tables \ref{table:loss_fullhessian} and
\ref{table:loss_upperbound}. Table \ref{table:loss_fullhessian} refers
to comparisons between \code{oem()} using the full Hessian and other
methods and Table \ref{table:loss_upperbound} refers to comparisons
between \code{oem()} using the Hessian upper-bound. We can see that
\code{oem()} in both cases results in solutions which are generally more
precise than other methods. Interestingly, for the logistic regression
simulations unlike the linear regression simulations, we see that
\code{oem()} provides solutions which are dramatically more precise than
\code{ncvreg()} for the MCP. These results indicate that the
nonconvexity of the MCP has more serious consequences for logistic
regression in terms of algorithmic choices.

\begin{table}[b]
\center
\begin{tabular}{llccc}
\toprule\toprule
$n$ & $p$ & \pkg{glmnet} & \pkg{gglasso} & \multicolumn{1}{c}{\pkg{ncvreg}} \\ 
\midrule
\nopagebreak $10^5$ & \nopagebreak 50  & $-4.50^{-09}$ & $-3.46^{-06}$ & $ \hphantom{-}2.47^{-11}$ \\
 & \nopagebreak 100  & $-5.38^{-09}$ & $-3.38^{-06}$ & $-1.46^{-12}$ \\
 & \nopagebreak 250  & $-1.93^{-08}$ & $-3.37^{-06}$ & $-1.15^{-12}$ \\
 & \nopagebreak 500  & $-3.58^{-08}$ & $-3.29^{-06}$ & $ \hphantom{-}7.11^{-12}$ \\
 \rule{0pt}{1.7\normalbaselineskip}
\nopagebreak $10^6$ & \nopagebreak 50  & $-2.00^{-09}$ & $-3.32^{-06}$ & $ \hphantom{-}7.00^{-14}$ \\
 & \nopagebreak 100  & $-4.34^{-09}$ & $-3.30^{-06}$ & $-1.39^{-14}$ \\
 & \nopagebreak 250  & $-8.82^{-09}$ & $-3.36^{-06}$ & $ \hphantom{-}6.76^{-12}$ \\
 & \nopagebreak 500  & $-1.75^{-08}$ & $-3.34^{-06}$ & $ \hphantom{-}1.69^{-14}$ \\
\bottomrule 
\end{tabular}
    \caption{The results above are the averages differences between the \code{oem()} function and other methods in objective function values averaged over all of the values of the tuning parameter for the linear regression simulations. Negative here means the \code{oem()} function results in estimates with a lower objective function value.}  
   \label{table:loss_linear}  
\end{table}

\begin{table}[b]
\center
\begin{tabular}{llcccc}
\toprule\toprule
$n$ & $p$ & \pkg{glmnet} & \pkg{glmnet} (ub) & \pkg{gglasso} & \multicolumn{1}{c}{\pkg{ncvreg}} \\ 
\midrule
\nopagebreak $10^4$ & \nopagebreak 50  & $-2.79^{-13}$ & $-4.30^{-12}$ & $-4.33^{-07}$ & $-5.31^{-03}$ \\
 & \nopagebreak 100  & $-1.82^{-13}$ & $-4.74^{-12}$ & $-4.95^{-07}$ & $-4.90^{-03}$ \\
 & \nopagebreak 250  & $-6.52^{-13}$ & $-3.87^{-12}$ & $-5.06^{-07}$ & $-5.03^{-03}$ \\
 & \nopagebreak 500  & $-4.80^{-13}$ & $-6.90^{-12}$ & $-5.30^{-07}$ & $-5.48^{-03}$ \\
 \rule{0pt}{1.7\normalbaselineskip}
\nopagebreak $10^5$ & \nopagebreak 50  & $-1.48^{-13}$ & $-2.72^{-12}$ & $-4.79^{-07}$ & $-3.05^{-03}$ \\
 & \nopagebreak 100  & $-1.30^{-13}$ & $-2.17^{-12}$ & $-4.78^{-07}$ & $-4.55^{-03}$ \\
 & \nopagebreak 250  & $-9.30^{-14}$ & $-2.78^{-12}$ & $-4.78^{-07}$ & $-4.47^{-03}$ \\
 & \nopagebreak 500  & $-2.07^{-13}$ & $-4.49^{-12}$ & $-4.51^{-07}$ & $-3.07^{-03}$ \\
\midrule 
\end{tabular}
    \caption{The results above are the averages differences between the full Hessian version of \code{oem()} and other methods in objective function values averaged over all of the values of the tuning parameter for the logistic regression simulations. Negative here means the \code{oem()} function results in estimates with a lower objective function value. The heading ``\pkg{glmnet} (ub)'' corresponds to \code{glmnet()} with the Hessian upper bound option.}  
   \label{table:loss_fullhessian}  
\end{table}

\begin{table}[b]
\center
\begin{tabular}{llcccc}
\toprule\toprule
$n$ & $p$ & \pkg{glmnet} & \pkg{glmnet} (ub) & \pkg{gglasso} & \multicolumn{1}{c}{\pkg{ncvreg}} \\ 
\midrule
\nopagebreak $10^4$ & \nopagebreak 50  & $-7.01^{-14}$ & $-4.09^{-12}$ & $-4.20^{-07}$ & $-5.31^{-03}$ \\
 & \nopagebreak 100  & $ \hphantom{-}6.74^{-13}$ & $-3.88^{-12}$ & $-4.76^{-07}$ & $-4.90^{-03}$ \\
 & \nopagebreak 250  & $-3.88^{-14}$ & $-3.25^{-12}$ & $-4.97^{-07}$ & $-5.03^{-03}$ \\
 & \nopagebreak 500  & $-4.06^{-13}$ & $-6.82^{-12}$ & $-5.16^{-07}$ & $-5.47^{-03}$ \\
 \rule{0pt}{1.7\normalbaselineskip}
\nopagebreak $10^5$ & \nopagebreak 50  & $ \hphantom{-}3.16^{-12}$ & $ \hphantom{-}5.87^{-13}$ & $-4.71^{-07}$ & $-3.05^{-03}$ \\
 & \nopagebreak 100  & $ \hphantom{-}3.06^{-12}$ & $ \hphantom{-}1.02^{-12}$ & $-4.61^{-07}$ & $-4.55^{-03}$ \\
 & \nopagebreak 250  & $ \hphantom{-}2.72^{-12}$ & $ \hphantom{-}3.40^{-14}$ & $-4.71^{-07}$ & $-4.47^{-03}$ \\
 & \nopagebreak 500  & $ \hphantom{-}1.09^{-12}$ & $-3.19^{-12}$ & $-4.31^{-07}$ & $-3.07^{-03}$ \\
\bottomrule 
\end{tabular}
    \caption{The results above are the averages differences between the upper bound version of \code{oem()} and other methods in objective function values averaged over all of the values of the tuning parameter for the logistic regression simulations. Negative here means the \code{oem()} function results in estimates with a lower objective function value. The heading ``\pkg{glmnet} (ub)'' corresponds to \code{glmnet()} with the Hessian upper bound option.}  
   \label{table:loss_upperbound}  
\end{table}

\section[Acknowledgments]{Acknowledgments}\label{acknowledgments}

This material is based upon work supported by, or in part by, the U. S.
Army Research Laboratory and the U. S. Army Research Office under
contract/grant number W911NF1510156, NSF Grants DMS 1055214 and DMS
1564376, and NIH grant T32HL083806.

\bibliographystyle{jss}\bibliography{references}

\end{document}